\documentclass[preprint,12pt]{article}
\usepackage{graphicx}

\title{Mutations as Levy flights}

\author{Dario A. Leon (1,2,*) and Augusto Gonzalez (3,2)\\
(1) University of Modena \& Reggio Emilia, 41125, Modena, Italy\\
(2) Institute of Cybernetics, Mathematics and Physics\\
 10400, Havana, Cuba\\
(3) University of Electronic Science and Technology\\ 
 610051 Chengdu, People Republic of China\\
(*) dario@icimaf.cu}

\begin{document}
\maketitle

\begin{abstract}
Data from a long time evolution experiment with Escherichia Coli and from a large study
on copy number variations in subjects with European ancestry are analyzed in order to argue that
mutations can be described as Levy flights in the mutation space. 
These Levy flights have at least two components: random single-base substitutions and large DNA
rearrangements. From the data, we get estimations for the time rates of both events and the
size distribution function of large rearrangements. 
\end{abstract}

\flushbottom


\section*{Introduction}
Life is coded in the DNA molecule, and the combined effect of random mutations and natural selection
leads to biological evolution. Point mutations provide a kind of fine tuning of the genome,
allowing the optimization of protein functions, for example. On the other hand, radical
remodeling by genetic recombination events is thought to be the source of global changes, 
leading even to new biological species \cite{MBC}.

In order to describe mutations, one shall determine the rate at which they occur and the ``spatial''
distribution function, that is their distribution along the DNA. To the best of our 
knowledge, there are precise measurements of the mutation rates in many situations \cite{rates1,
rates2,rates3}, as well as precise indications of sites or regions in the DNA prone to mutations 
\cite{hotspots}. However, there are no results concerning the length distribution function of 
mutations.

In the present paper, we use data from a long-term evolution experiment (LTEE) with E. Coli
populations \cite{Lenski} in order to get the rate of both, point mutations and large
chromosomal rearrangement events in the evolution of this bacterium. 
Data on single-nucleotide polymorphisms (SNPs) in mixed-population samples, taken from 
generation 2000 to 40000, come from sequencing these samples and aligning to the genome sequence
of the ancestral strain \cite{SPM}. On the other hand, large chromosomal rearrangements
in clones harvested from these samples are identified by means of a combination of optical
techniques, genome sequencing and PCR analysis \cite{LR}. The emergence of a mutator phenotype, which increases the rate of SPMs by 100 times but does not affect large rearrangements, indicates that these are essentially different processes.
However, although there are many types of large rearrangements, responding to different scales and  mechanisms, all of them can be accommodated into a  global distribution function  for the lengths of the modified DNA segments, exhibiting a scale-free power-like behavior.

On the other hand, the rate of single point mutations (SPMs) and other DNA 
rearrangements in germline cells in humans has been precisely measured \cite{rates3,rates4}. 
With regard to the length distribution function of mutations, we shall use data from a 
recent large study on copy number variants in subjects with European ancestry \cite{cnv}.
As in the bacterial case, a scale-free distribution function arises in the scale range 
spanned by the experiment. In both, bacteria and human germline cells, we show that the fit can be extended to the small -length range of the data by means of a stable Levy distribution  
\cite{Levy1}.

From an abstract point of view, mutations can be described as a succession of
transformations in the DNA molecule - a Markov chain \cite{markov}. The chain configuration at the step $i+1$ is written in terms of the configuration at step $i$ as: $X_{i+1}=X_i+\delta_i$, where $\delta_i$ is the introduced modification. The main result of the paper is that in $\delta_i$ we shall distinguish at least two kinds of transformations: SPMs and large rearrangements, the lengths of the latter are distributed 
according to a stable Levy law. That is, mutations are a kind of Levy flights 
\cite{Levy2}.
\vspace{.5cm}

\section*{Data on bacterial SPMs}
\begin{figure}[ht]
\begin{center}
\includegraphics[width=0.8\linewidth,angle=0]{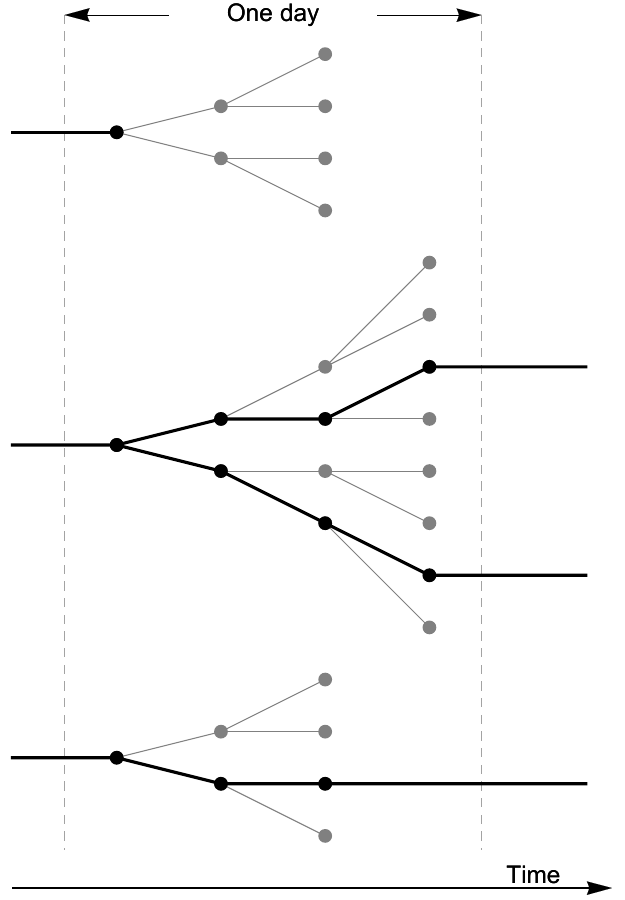}
\caption{Phylogenetic representation of one day evolution in the LTEE. After a few clonal 
divisions (2-3 in the figure, 6-7 in the experiment) individuals are randomly selected to pass
to the next day. Most lineages are truncated, whereas those with higher fitness have better
possibilities to continue to the next day.}
\label{fig1}
\end{center}
\end{figure}

In an evolution experiment, random fluctuations are filtered by natural selection. The
evolution dynamics in the LTEE is schematically represented in Fig. \ref{fig1}. Cell
lineages with neutral or deleterious mutations are usually truncated, whereas beneficial
mutations confer evolutionary advantage to clones and, thus, higher probability to continue.
Once they appear, beneficial mutations are fixed in more than 50 \% of the population
after a fixing time. The number of cell lineages, that is number of cells passing to the next day in the evolution, is kept fixed to around five millions in the experiment.

\begin{figure}[ht]
\begin{center}
\includegraphics[width=0.9\linewidth,angle=0]{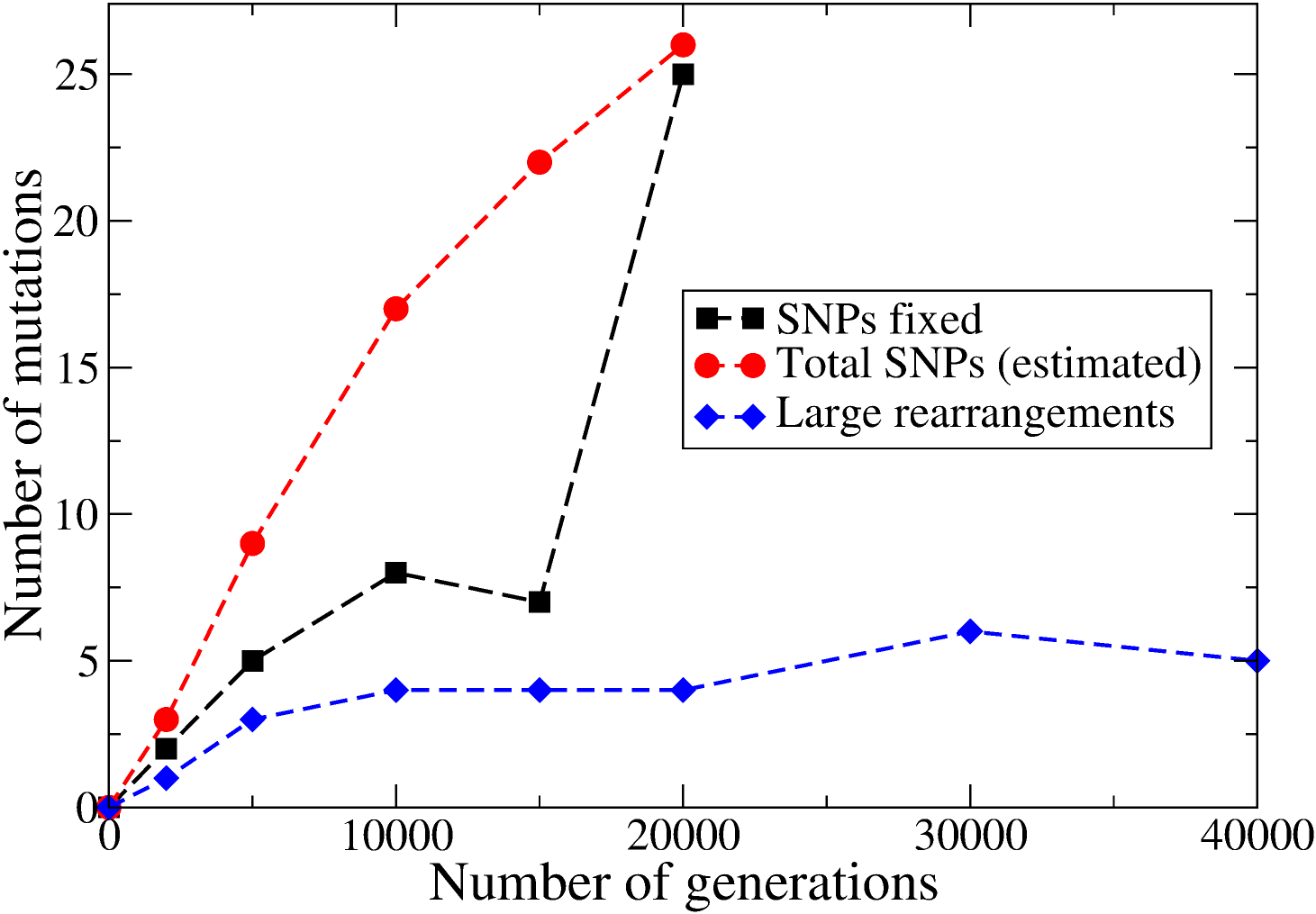}
\caption{The number of observed mutations as a function of time (number of generations) in a population named
Ara-1 of the LTEE \cite{SPM}. Data from generation 0 (ancestral strain, taken as reference) 
to 20000 are included in the figure. Fixed SNPs (black squares), the estimated mean number of SNPs
in clones (red dots, coming from calculations in the Appendix), and the number of large rearrangements (blue diamonds) are shown.}
\label{fig2}
\end{center}
\end{figure}

We draw in Fig. \ref{fig2} the data on SPMs, taken from Ref. \cite{SPM}. A population, 
called Ara-1 in the experiment, is sampled at generations 2000, 5000, 10000, 15000, 20000, 30000, and
40000. The two latter points are not included in the figure because of a mutator phenotype, which
appears at generation 27000 and leads to a 100-fold increase of the mutation rate.

Alignment of 36-base reads in mixed population samples yielded 40- to 60-fold coverage, allowing
to determine frequencies of SPMs above 4 \% in the population. Authors report ``fixed'' SPMs,
meaning that their frequency, $f$, is above 96 \%, as well as so called 
single-nucleotide polymorphisms, SNPs, where 4 \% $< f <$  96 \%.

The data labeled ``fixed'' in the figure, corresponding to mutations with $f\ge 96$ \%,
show a linear increase at short times with a slope $1.0\times 10^{-3}$ mutations/generation. 
The data labeled ``mean'', on the other 
hand, correspond to our estimation for the mean number of mutations one may detect in a clone 
(see Supplementary Material for details). The slope of the mean curve at short times is a little higher,
around $1.8\times 10^{-3}$ mutations/generation,
which may be taken as an estimation of the SPM rate, $p_{SPM}$. 

The value obtained for $p_{SPM}$ should be compared with the total point mutation rate, 
that in the LTEE was estimated to be 
$p_{SPM}=10^{-4}$ - $10^{-3}$ mutations per generation for the whole genome
\cite{bareSPM}. 

These are effective rates. A detailed simulation of the evolutionary dynamics 
requires including competition between clones, drift processes, etc \cite{epistasis,dynamics}.




Summarizing the present section, we may say that, for SPMs along a cell lineage, we estimate 
$p_{SPM}\sim 1.8\times 10^{-3}$ mutations per generation.
The latter is obtained from the slope of the model curve in Fig. \ref{fig2} near the origin.

Finally, a very significant point is that, in the studied population, a mutator genotype 
becoming dominant after generation 27000 increases by a factor of 100 the point mutation rate. 

The next section is devoted to large rearrangements. 
\vspace{.5cm}

\section*{Data on large chromosomal rearrangements in the LTEE}
Data on large chromosomal rearrangements are provided in Ref. \cite{LR}. Due to 
experimental limitations, authors can not reliably detect rearrangements smaller than
5 Kilo base pairs (Kbp). On the other hand, they perform measurements on clones, 
that is representatives of a population, which may exhibit strong deviations from mean
values.

Let us stress that mutations are rare events. The data reported in Refs. \cite{SPM} and \cite{LR}
are the results of 20 years of evolution and 40000 bacterial generations.
However, only around 100 large chromosomal rearrangements are registered in the
12 populations under study. With such scarce 
data we can not pretend a precise description of the mutation distribution function. 
Only qualitative and semi-quantitative results can be extracted. 

The first set of results involve a time sequence of clones of the population Ara-1, as
in the previous section. That is, samples at generations 2000, 5000, 10000, 15000, 20000, 
30000, 40000, and 50000. 


We shall estimate the time rate, $p_{LR}$, and size probability
distribution, $\pi_{LR}(l)$, of such events. The experiments report on different kinds 
of rearrangements: deletions, insertions, translocations, and inversions. 

We included in Fig. \ref{fig2} the detected number of large rearrangement events as a function of time 
(number of generations). Half of these rearrangements seem to be fixed, in the sense 
that they are detected also at later times in different clones. 
From the slope at short times, we get a rough estimation for the rate of large changes, 
$p_{LR}\sim 5\times 10^{-4}$ large chromosomal rearrangements/generation, a value three times smaller
than $p_{SPM}$. 

Because of the experimental resolution, not all the rearrangements are registered, specially 
short-length ones. Thus, our estimation for $p_{LR}$ is a lower bound, and the actual rate could be 
similar to $p_{SPM}$.

A second very important point is related to the fact that this figure does not show 
any abrupt increase of $p_{LR}$ after generation 27000,
where the mutator phenotype becomes dominant. This fact stresses the differences 
between the mechanisms leading to SPMs and large rearrangements in the bacterial
chromosome.

\begin{figure}[ht]
\begin{center}
\includegraphics[width=0.7\linewidth,angle=0]{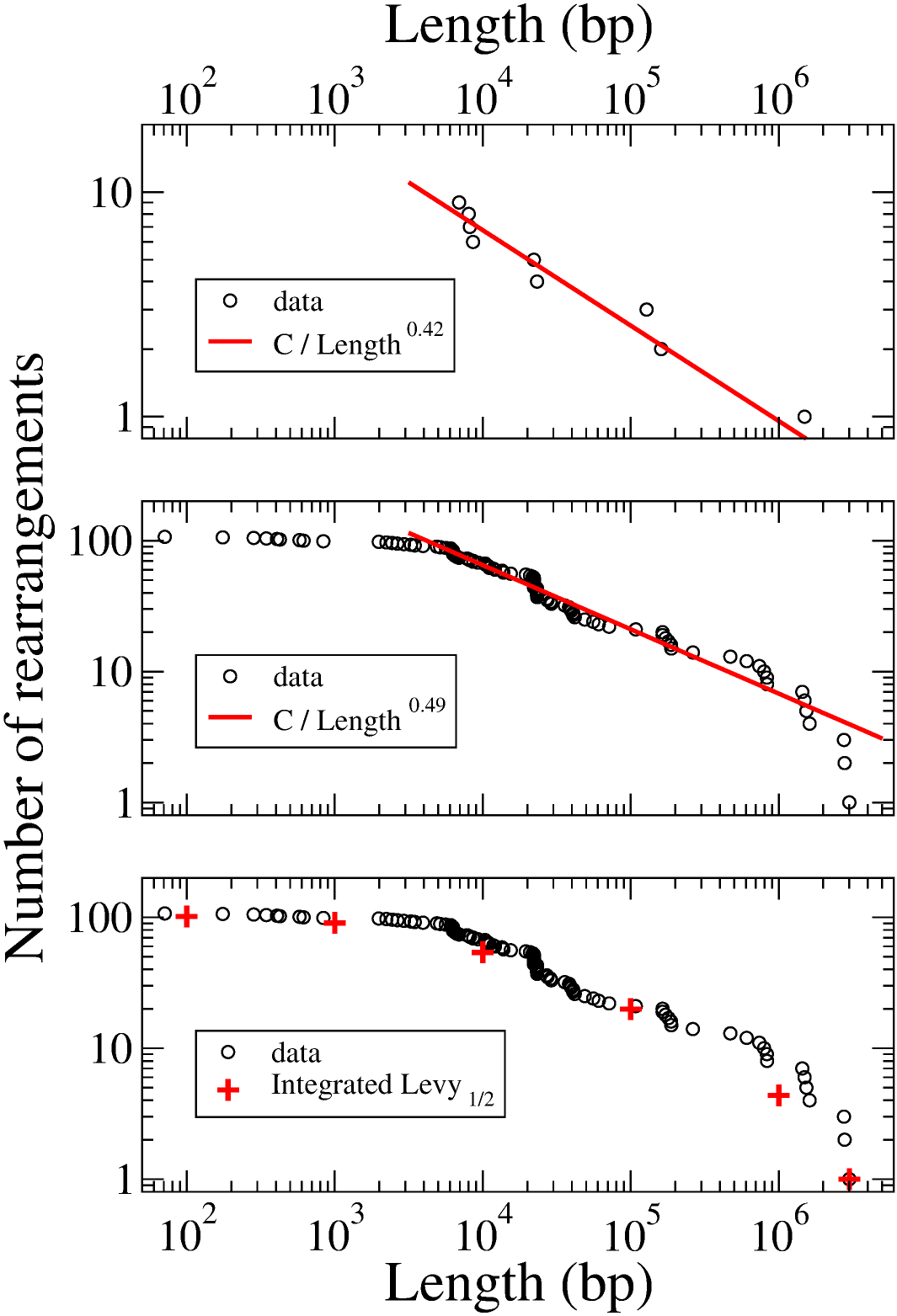}
\caption{Top: Log-log plot of the integrated size distribution function of large (greater than 5 Kbp) chromosomal rearrangements in clones of the Ara-1 population. The red line is a  fit with a $1/l^\nu$ dependence. Center: Log-log plot of the integrated size distribution of large rearrangements in clones obtained from the 12 independently evolving populations in the LTEE, sampled at generation 40000. The red line is a fit with the function $1/l^\nu$ for $l$ in the interval $5 \times 10^3 < l < 1.7 \times 10^6$. Bottom: A fit to the observed distribution in a wider interval by means of the integral Levy 1/2 distribution given in Eq. (\ref{eq3}).}
\label{fig3}
\end{center}
\end{figure}

Fig \ref{fig3}, upper panel, on the other hand, reflects the size statistics. We use 
a log-log plot. The $x$-axis is the size, $l$, and the $y$-axis is the number of 
rearrangements with size greater or equal than $l$. In the interval $5\times 10^3 
< l < 1.5\times 10^6$, the data are very well fitted by the function 
$C/l^{\nu}$, with $\nu= 0.42$ and $C$ a normalization constant (Pearson
correlation coefficient $r=0.96$).

This dependence can be understood as coming from a probability $\sim \nu/l^{1+\nu}$
for a large rearrangement of size $l$ to occur. Indeed, the number of events
with size greater or equal than $l$ is thus computed as:

\begin{equation}
C \nu \int_l^{\infty}\frac{{\rm d}x}{x^{1+\nu}}=\frac{C}{l^{\nu}}.
\end{equation}

These results are based on the 9 detected large DNA rearrangement events in the
Ara-1 population in 50000 generations.
Below, we shall consider a larger data with better statistics. 
The data comes from clones harvested from the 12 independently 
evolving populations in the LTEE, sampled at generation 40000. There are 110 detected 
large rearrangements in these clones. The results are shown in Fig. \ref{fig3} center panel.

First, we shall stress that the points for $5\times 10^3 < l < 1.7\times 10^6$ show a
remarkable scaling with $\nu= 0.49$ (Pearson correlation coefficient $r=0.99$). 
The slope of the experimental curve changes 
for $l<5$ Kbp. This fact may be partially due to the limitations of the experimental
techniques that can not detect all of the rearrangements 
for these $l$ values, as mentioned by the authors. But it could also be related to
a saturation of the distribution function for small values of $l$. 
On the other hand, in the right hand side of the figure $l$ is near the bacterial DNA 
size, $L\approx 5\times 10^6$ bp.

The idea behind this figure is to show that the length distribution function of all the evolving populations is similar. A common exponent near 1/2 seems to describe all the populations.

Up to this point we have concluded that SPMs and large rearrangements have essentially different mechanisms for their time rates, but the many types of large rearrangements can be accommodated into a common distribution function. For this latter property to hold, the distribution function should obey the central limit theorem, that is the sum of independent sub-processes should preserve the function. 

Motivated by these facts, that is the central limit theorem and saturation in the low-length region, we tried a stable Levy distribution $L_{1/2} (\alpha l)$ in order to 
fit the observed distribution of points. In general \cite{Levy1}, the Levy probability
density distribution

\begin{equation}
L_{\nu}(y)=\frac{2}{\pi}\int_0^{\infty}\exp(-q^\nu)\cos(q y){\rm d}q
\end{equation}

\noindent
behaves as $1/y^{\nu+1}$ for large values of its argument.

In Fig. \ref{fig3} bottom panel, the integrated probability density is plotted. As the length 
runs from the maximal value, $l_{max}$, to its minimum, $l_{min}$, the number of
rearrangements rises from 1 to the total value, $N_{LR}$. The integrated distribution, 
red crosses in Fig. \ref{fig3} bottom panel, may thus be written as:

\begin{equation}
f(l)=1+(N_{LR}-1)\frac{\int_l^{l_{max}} L_{1/2}(\alpha y)~{\rm d}y}
 {\int_{l_{min}}^{l_{max}} L_{1/2}(\alpha y)~{\rm d}y}.
\label{eq3}
\end{equation}

\noindent
The parameter $\alpha=10^{-4}$ provides a very good fit. 

Summarizing the section, we may say that large rearrangements are
observed in bacterial cell lineages at rates $p_{LR}\sim 5\times 10^{-4}$ 
per generation. This is a lower bound for $p_{LR}$, the actual value could be closer to 
$p_{SPM}$. No changes in $p_{LR}$ are reported after generation 27000
in the Ara-1 population, when a mutator
phenotype leads to a 100-fold increase of $p_{SPM}$, which means that the mechanisms 
responsible for SPMs and LRs are very different. The observed rearrangements are
well described by a stable Levy distribution $L_{1/2}(\alpha l)$, which in the reliable 
size interval, $5\times 10^3 < l < 1.7\times 10^6$, shows a dependence $\sim 1/l^{3/2}$.
\vspace{.5cm}

\section*{The rate of mutations in human germline cells}
We shall consider mutations in the human germline cells. Somatic mutations, although
relevant in aging processes, cancer, etc are less constrained by evolution and may be
dictated by different rules.

The natural unit of time in the present case, instead of cell generations, are organism
generations, that is births. The data is summarized in Fig. 1 of paper \cite{rates4},
where single nucleotide variants (SNVs) are distinguished from rearrangements such as 
small indels (mean length 2 bp), mobile elements insertions (MEIs, mean length 200 bp),
copy number variants (CNVs, mean length $10^6$ bp), and aneuploidies (mean length $10^8$ bp). 
There are around 60 SNVs per birth, five times the number of all other mutations taken 
together. 

Although there is not a complete understanding of mechanisms causing these kinds of
mutations, one should expect different acting mechanisms and, thus, independent random
processes. In a model of mutations we shall consider, as in bacteria, at least two independent 
processes: SNVs and chromosome rearrangements. The former, of Brownian character, acting with a 
rate of 60 mutations per birth; and the latter, with a rate of around 10 mutations per birth
and a length distribution function which shall be determined. In the next section it will become 
apparent that CNV events can be described by a stable Levy function.
\vspace{.5cm}

\section*{The length distribution function of CNVs}
We use data from a recent study of CNVs in more than 100000 subjects of European ancestry \cite{cnv}. 
Typical CNVs have lengths below 1 Mbp and frequencies below 0.01 in the
studied cohort. However, the authors provide data for more than 1.7 millions of rearrangements 
which lengths range from 10 to $2.4 \times 10^8$  bp. Not all mutations are detected with the same fidelity
in this wide range. One should expect short-length mutations to be under counted.

The low frequencies of mutations indicate that they are mostly of neutral or deleterious character.

\begin{figure}[ht]
\begin{center}
\includegraphics[width=0.9\linewidth,angle=0]{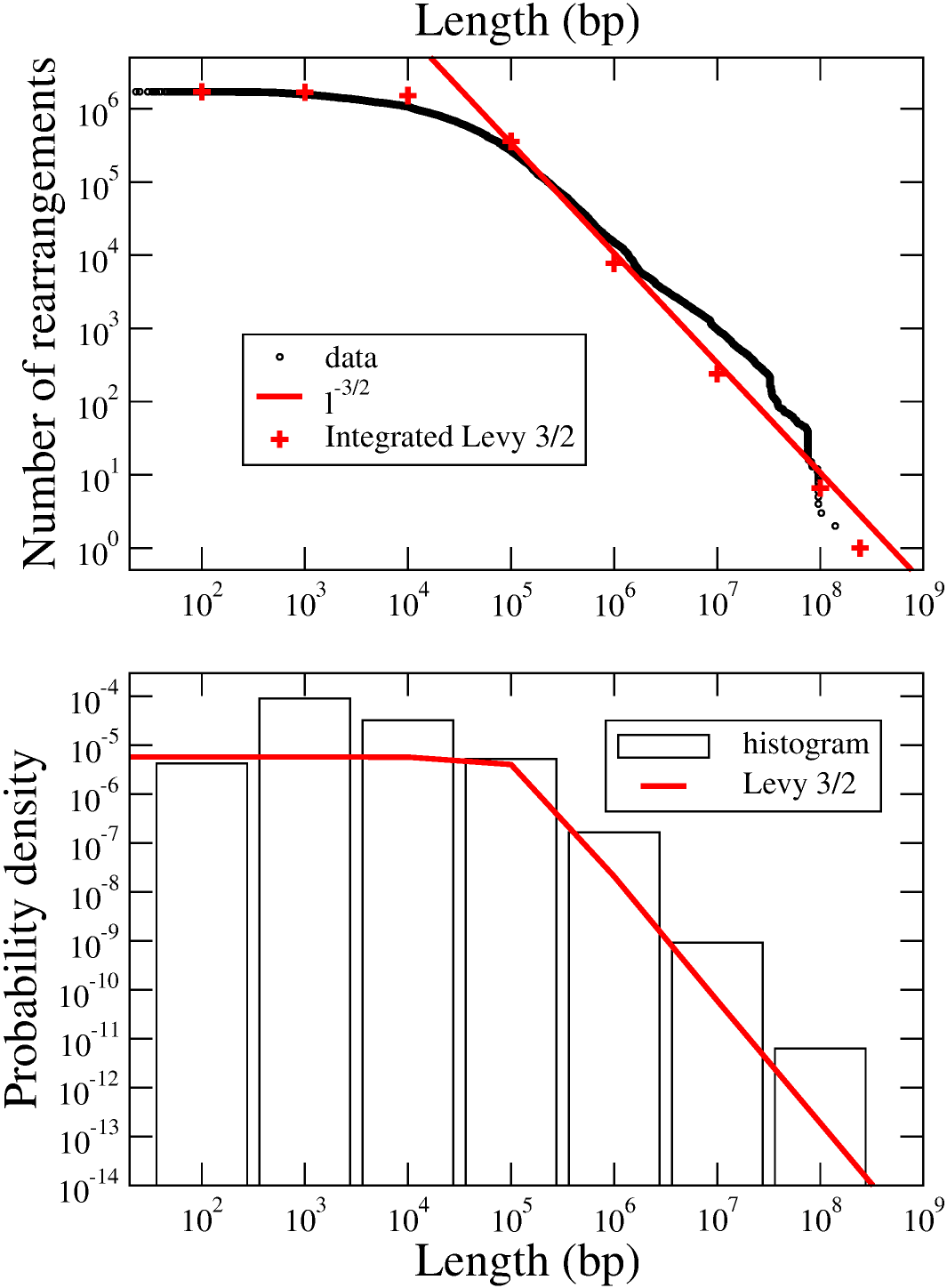}
\caption{Top: Log-log plot of the integrated size distribution of CNVs in germline cells.
The red line is a fit with the function $1/l^{3/2}$ for $l > 10^5$ bp whereas the red crosses come
from the integral Levy 3/2 distribution. Bottom: A direct comparison of $\alpha L_{3/2}(\alpha l)$
with a log-contracted histogram of the data.}
\label{fig4}
\end{center}
\end{figure}

The results are presented in Fig. \ref{fig4} top panel. This figure is similar to Fig. \ref{fig3} for
bacteria. The $x$ axis is the length, $l$, of the mutated segment, and the $y$ axis is the number
of rearrangements with lengths greater than or equal to $l$. Thus, this is an integrated probability distribution
and we expect it to be described by a formula like Eq. (\ref{eq3}), in which $L_{1/2}$ is replaced by
$L_{\nu}$, and the parameters $l_{max}$, $l_{min}$ and $N_{LR}$ are actualized accordingly.

The figure shows that the distribution can be well fitted by a $\nu=3/2$ stable Levy
function, $L_{3/2}(\alpha l)$, with a scale parameter $\alpha=10^{-5}$. A
tail $\sim l^{-3/2}$ is apparent in the integrated distribution function for lengths
greater than $1/\alpha\sim 10^5$ bp.

The bottom panel of this figure contains a direct comparison between the 
$\alpha L_{3/2}(\alpha l)$ probability density and a histogram in which the $x-$axis
is log contracted such that each bin spans a decade. Fluctuations are apparent in the figure,
specially in the short-length region, where mutations are most likely to be undetected.
This is the reason why we decided to fit the smoother integrated probability function, instead 
of the probability density. The exponent $\nu\approx 3/2$ seems to be a robust determination.

To summarize the section we may state that a model of mutations in human germline cells 
should contain at least two processes: Brownian SNVs with a rate of around 60 mutations per birth, 
and chromosome rearrangements with a stable Levy 3/2 distribution function for the lengths,
which become scale-free for $l>10^5$ bp. The rate of the latter events is probably well below 
10 events per birth. Additional short- and intermediate-lengths measurements should be
conducted in order to precise whether they can be described merely by a modification of
the scale $\alpha$ in the Levy 3/2 function or should be included as independent random processes.   
\vspace{.5cm}

\section*{Discussion}
The data on SPMs and large rearrangements in bacterial DNA in the course
of 50000 generations of evolution seem to support a picture in which 
both kinds of events occurs with similar rates \cite{tempo}. 
The size distribution of large rearrangements can be fitted with
a stable Levy function with exponent $\nu\approx 1/2$ and scale 
$\alpha\approx 10^{-4}$. 

This is a kind of Levy flight picture for mutations along a cell
lineage in which small deviations and radical changes in the genome 
are combined. 

In a way, our paper is similar to Refs. \cite{foraging1,foraging2}, where the Levy flight
theory of foraging is tested against experimental data.

The picture is not complete, however, because of the lack 
of experimental data on chromosomal rearrangement, in the 
range $1<l<5\times 10^3$ bp. Notice that the inverse of $\alpha$ coincides with 
both the lower range of experimental observations and the length
above which the integrated distribution reaches its asymptotic behavior
$\sim 1/l^{\nu}$. 

For $l > 1/\alpha$ there are no additional scales, and
the distribution function is roughly  scale-free.
This is somehow unexpected. Naively, one would expect a scale of the order 
of a few Kbp and a rapidly decaying distribution function for rearrangement lengths 
larger than the scale. The biological mechanism by which such a 
scale-free distribution is generated should be further clarified.

We stress that a power-like scale-free distribution is observed also for the distances (spacers) between highly conserved fragments in several genomes  \cite{plos}. 
We guess that an evolutionary model  for mutations in which Levy flights are constrained to respect conserved fragments would lead to a power-like distribution
function for distances between fragments.

The LTEE is a clean clonal evolution experiment. In wild conditions, horizontal gene transfer through  recombination events is expected to play an important role \cite{ref1}. 
The lengths of recombined fragments seem to be distributed along a power law also \cite{ref2}. 

From an abstract perspective, a scale-free distribution for large 
rearrangements is a good strategy \cite{foraging1}. In the described experiment, where the population 
size is controlled and nutrients are limited, biological evolution can be 
viewed as an optimization problem. 
The mean fitness in the population is the cost function. Mutations provide the mechanism for
surveying the parameter space, and natural selection picks up the best representatives
in the population. A local search alone, like the SPMs or short length rearrangements, 
could trap mutation trajectories around a local maximum in the fitness landscape. 
An optimal search algorithm shall include large rearrangements of any size, that 
is a scale-free size distribution.

The near optimal character of the search algorithm is confirmed in the experiment by 
what authors call ``parallel mutations'' \cite{LR}, that is very similar fixed 
mutations in independently evolving populations.

We notice, by the way, that the idea of a Levy search has been implemented in computational 
optimization techniques \cite{Levy1}. 

We also checked our statement about the Levy nature of mutations 
in eukaryotes, in particular in human germline cells. Recent precise data on CNVs
allowed the determination of the length distribution function in scales larger than
$10^4 - 10^5$ bp. We could fit the distribution to a stable Levy 3/2 function,
which shows a scale-free behavior up to the typical chromosome length.
However, in the short-length region, related to small indels and MEIs, 
one expects that the data is incomplete,
and we can not distinguish whether this lower scale region can be described simply by a modification
of the parameter $\alpha$ or independent random processes should be included in the theory.

The fact that the biological complex processes leading to mutations, 
probably originated from many different mechanisms,
exhibit scaling in a very wide range of lengths should be based on very general laws.  
Our idea to use a stable Levy function in order to fit the data is
motivated by such arguments. Stable functions, respecting the central limit theorem,
are very good candidates.

We notice that very general arguments have been suggested to explain the observed
power-like (Pareto) distribution function for gene expression \cite{GE2} in cells. 
Our paper is similar in spirit  to this one.

Differently from the conclusions of the LTEE experiment, the low frequencies observed in CNVs
indicate that most of these mutations exhibit neutral or deleterious character, and indeed they 
are shown to be strongly correlated to
diseases or disorders \cite{cnv}.

The data, although limited, seems to suggest exponents 1/2 and 3/2 for bacteria and human germinal cells, respectively. 
New questions arise as, for example, whether the exponents, and not only the mutation rates, may vary under different 
selective pressure, whether the change from 1/2 to 3/2 reflects a trend in evolution \cite{evo}, etc. 
On the other hand, it is known that the optimal value for the exponent in Levy searches is equal to one \cite{LevyPNAS1}. 
The obtained exponents are close to this value. There are also known limitations of Levy searches, in particular 
to find close minima \cite{LevyPNAS}.  
The question arise as to whether the addition of SPMs as an independent process in our model (and, probably, other short-length processes) is a way of correcting such limitations.  

A probable next step in our research would be to describe the very important somatic mutations, 
involved in aging processes and cancer. 
Somatic stem cells in human tissues have been shown to reach numbers above $10^8$,
and their replication rates may lead to $10^4$ cell generations along a 
lifespan \cite{Tomasetti}, a number comparable to the number of generations
in the controlled LTEE with E. Coli. 

Massive sequencing of tumors are already available, 
see for example \cite{sequencing}, and the importance of somatic mutations in cancer
is widely recognized. A catalogue of somatic mutations in cancer exists\\ 
(https://cancer.sanger.ac.uk/cosmic), which may provide the data for checking the Levy hypothesis.
The idea that large rearrangement hits on particular genes may lead to cancer
is very plausible. In particular, hits on very important genes, such as p53 \cite{p53}.
Correlations between CNVs and relevant genes have been tested \cite{cnv}.

If the Levy nature of mutations is generally confirmed, it could have practical 
implications in modeling carcinogenesis. 
The key obstacle is to relate mutations to cellular fitness \cite{ref1}. In gene expression space \cite{cancer}, however, the high and low fitness regions are apparent. Normal tissues and tumors are grouped in disjoint high fitness regions. We have tried\cite{cancer} a local plus Levy jumps model for the motion in this space that seems to reproduce the data on cancer risk in a set of tissues.
\vspace{.5cm}










\appendix
\section*{Appendix. Estimated mean number of SPMs in a clone}
\label{app}
Let us consider a mixed bacterial population, where there are $N_{fixed}$ fixed 
mutations (frequencies $\ge 96 \%$), and a number of additional SNPs with 
frequencies $f_i$. We assume that these mutations are not correlated, that is a 
given frequency $f_i$ is independent from any other $f_j$. The probability of finding one additional mutation in a clone is, thus:

$$ P_{1}= \sum_i f_i.$$ 

If $P(1)> 0.5$ we say that the mean number of mutations 
we may find in a clone is, at least, $N_{fixed}+1$.

Similarly, we define:

$$ P(2)=\sum_{i<j} f_i f_j,$$ 

\noindent
and state that the mean number of mutations is, at least, $N_{fixed}+2$ if
$P(2)> 0.5$.

The probabilities for $N_{fixed}+3$, $N_{fixed}+4$, etc mutations are 
defined in the same way.

We shall say that the mean number of mutations we may find in a clone is
$N_{fixed}+n$ if $P(n)> 0.5$, but $P(n+1)< 0.5$.

In order to draw the ``model'' curve in Fig. 2 of the main manuscript, we use the data of paper
\cite{SPM} for the frequencies of observed SNPs and compute the mean number
of mutations in clones.

\bibliography{sample}
\bibliographystyle{unsrt}

\section*{ Acknowledgments}
A.G. acknowledges the Cuban Program for Basic Sciences, the Office of External Activities of 
the Abdus Salam Centre for Theoretical Physics, and the University of Electronic Science and 
Technology of China for support. The research is carried on under a  project of the Platform 
for Bio-informatics of BioCubaFarma, Cuba. Authors are grateful to the referees for comments and suggestions.

\section*{Author's contributions}
A.G. conceived and coordinated the work. D.A.L. and A.G. processed the experimental data
on bacteria. A.G. processed the experimental data on humans. D.A.L. is in charge of the GitHub repository. Both authors analyzed and 
interpreted the results, contributed to the manuscript and approved the final version.

\section*{Competing interests}
  The authors declare that they have no competing interests.

\section*{Availability of data and materials}

	The information about the data we used, the procedures and results are integrated in a public repository that is part of the project "Processing and Analyzing Mutations and Gene Expression Data in Different Systems": https://github.com/DarioALeonValido/evolp.
	
	The data we use for bacteria \cite{Lenski} and copy number variations \cite{cnv} are replicated in paths \verb+../evolp/bases_external/LTEE/mutations/+ \\ and \verb+../evolp/bases_external/CNV/+ respectively. To process each data set we include specific scripts in \verb+../evolp/Levy_mutations/+.


\end{document}